\documentclass[preprint,12pt]{elsarticle}

\usepackage{amssymb}

\begin{document}

\begin{frontmatter}



\title{Initial Results of a Silicon Sensor Irradiation Study for ILC Extreme Forward Calorimetry \\ \bigskip Talk presented at the International Workshop on Future Linear Colliders (LCWS13), Tokyo, Japan, 11-15 November 2013.}


\author[ucsc]{Reyer Band}
\author[ucsc]{Vitaliy Fadeyev}
\author[SLAC]{R. Clive Field}
\author[ucsc]{Spencer Key}
\author[ucsc]{Tae Sung Kim}
\author[SLAC]{Thomas Markiewicz}
\author[ucsc]{Forest Martinez-McKinney}
\author[SLAC]{Takashi Maruyama}
\author[ucsc]{Khilesh Mistry}
\author[ucsc]{Ravi Nidumolu}
\author[ucsc]{Bruce A. Schumm \corref{cor1}}
\ead{baschumm@ucsc.edu}
\author[ucsc]{Edwin Spencer}
\author[ucsc]{Conor Timlin}
\author[ucsc]{Max Wilder}
\cortext[cor1]{Corresponding author.}
\address[ucsc]{Santa Cruz Institute for Particle Physics and the University Of California, 1156 High Street,
Santa Cruz California 95064 USA}
\address[SLAC]{SLAC National Accelerator Laboratory, 2575 Sand Hill Road, Menlo Park California 94025 USA}

\address{}

\begin{abstract}

Detectors proposed for the International Linear Collider (ILC)
incorporate a tungsten sampling calorimeter (`BeamCal') intended to 
reconstruct showers of electrons, positrons and photons
that emerge from the interaction point of the collider
with angles between 5 and 50 milliradians. For the 
innermost radius of this calorimeter, radiation doses
at shower-max are expected to reach 100 MRad per year,
primarily due to minimum-ionizing electrons and positrons
that arise in the induced electromagnetic showers
of e+e- `beamstrahlung' pairs produced in the ILC beam-beam interaction. However,
radiation damage to calorimeter sensors may be dominated
by hadrons induced by nuclear interactions of shower photons,
which are much more likely to contribute to the non-ionizing
energy loss that has been observed to damage sensors exposed to
hadronic radiation. We report here on the results of SLAC
Experiment T-506, for which several different types of 
silicon diode sensors were exposed to doses of radiation
induced by showering electrons of energy 3.5-10.6 GeV. By embedding
the sensor under irradiation within a tungsten radiator, the exposure
incorporated hadronic species that would potentially contribute to the 
degradation of a sensor mounted in a precision sampling calorimeter. 
Depending on sensor technology, efficient charge collection
was observed for doses as large as 220 MRad.

\end{abstract}


\end{frontmatter}


\section{Introduction}
Far-forward calorimetry, covering the region between 5 and 50 milliradians
from the on-energy beam axis,
is envisioned as a component of both the ILD~\cite{ref:ILD_DBD} and SiD~\cite{ref:SiD_DBD}
detector concepts for the proposed International Linear Collider (ILC). The
BeamCal tungsten sampling calorimeter proposed to cover this angular region
is expected to absorb approximately 10 TeV of electromagnetic radiation
per beam crossing from e+e- beamstrahlung pairs, leading to expected annual radiation doses of 100 MRad
for the most heavily-irradiated portions of the instrument.
While the deposited energy is expected to arise primarily from minimum-ionizing
electrons and positrons in the induced electromagnetic showers, 
radiation damage to calorimeter sensors may be dominated
by hadrons induced by nuclear interactions of shower photons,
which are much more likely to contribute to the non-ionizing
energy loss that has been observed to damage sensors exposed to
hadronic radiation. We report here on the results of SLAC
Experiment T-506, for which several different types of
silicon diode sensors were exposed to doses of up to 220 MRad
at the approximate maxima of electromagnetic
showers induced in a tungsten radiator by electrons of energy
3.5-10.6 GeV, similar to that of electrons and positrons
from ILC beamstrahling pairs.

Bulk damage leading to the suppression of the electron/hole
charge-collection efficiency is generally thought to be proportional
to the non-ionizing energy loss (`NIEL') component of the energy
deposited by the incident radiation.
Early studies of electromagnetically-induced damage to
solar cells~\cite{ref:TaeSung5,ref:TaeSung6,ref:TaeSung7}
suggested that p-type bulk sensors were more tolerant to
damage from electromagnetic sources, due to an apparent
departure from NIEL scaling, particularly for electromagnetic particles
of lower incident energy. 

Several more-recent studies have 
explored radiation tolerance to incident fluxes of electrons.
A study assessing the capacitance vs. bias voltage (CV) characteristics of sensors exposed
to as much as 1 GRad of incident 2 MeV electrons~\cite{ref:Rafi_09}
suggested approximately 35 times less damage to n-type magnetic
Czochralski sensors than that expected from NIEL scaling.
A study of various n-type sensor types exposed to 900 MeV electrons 
showed charge-collection loss of as little as 3\% for exposures up to
50 MRad exposure~\cite{ref:Dittongo2004}; for exposures of 150 MRad, a suppression of damage
relative to NIEL expectations of up to a factor of four was observed~\cite{ref:Dittongo2005}. 
These discrepancies have been attributed to the different types of  
defects created by lattice interactions: electrons tend to create point-like defects that are more 
benign than the clusters formed due to hadronic interactions.

Finally, in studies of sensors exposed to large doses of hadron-induced
radiation, p-type bulk silicon was found to be more radiation-tolerant
than n-type bulk silicon, an observation that has been attributed to the absence of type inversion and the
collection of an electron-based signal~\cite{ref:TaeSung3,ref:TaeSung4}.
However, n-type bulk devices have certain advantages, such as a natural inter-electrode 
isolation with commonly used passivation materials such as silicon oxide and silicon nitride.

Here, we report on an exploration of the radiation tolerance of silicon sensors,
assessed via direct measurements of the median collected charge deposited 
by minimum-ionizing particles,
for four different bulk compositions: p-type and n-type doping of
both magnetic Czochralski and float-zone crystals.
The p-type float-zone sensors were produced by Hamamatsu Photonics while the
remaining types were produced by Micron Corporation.
Sensor strip pitch varied between
50 and 100 $\mu$m, while the bulk thickness 
varied between 307 $\mu$m (for the p-type magnetic Czochralski sensors) 
and 320 $\mu$m (for the p-type float zone sensors).
The use of these sensors is being explored as an alternative to several
more novel sensor technologies that are currently under
development~\cite{ref:ILD_DBD}, including GaAs and CVD diamond.

While the radiation dose was initiated by electromagnetic processes
(electrons showering in tungsten), the placement of the sensors near
shower max ensures that the shower incorporates an appropriate 
component of hadronic irradiation arising from neutron spallation,
photoproduction, and the excitation of the $\Delta$ resonance.
Particularly for the case that NIEL scaling suppresses 
electromagnetically-induced radiation damage, the small 
hadronic component of the electromagnetic
shower might dominate the rate of damage to the sensor.
However, the size and effect of this component is difficult to 
estimate reliably, and so we choose to study radiation damage in
a configuration that naturally incorporates all components present in
an electromagnetic shower.

\section{Experimental Setup}

Un-irradiated sensors were subjected to current vs. bias voltage (IV) and CV tests, 
the results of which allowed
a subset of them to be selected for irradiation based on their
breakdown voltage (typically above 1000 V for selected sensors) and low level of leakage 
current. The sensors were placed on carrier printed-circuit `daughter boards' and wire-bonded to a 
readout connector. The material of the daughter boards was milled away in the 
region to be irradiated in order to facilitate the
charge collection measurement (described below) and minimize radio-activation. 
The median collected charge was measured with the Santa Cruz Institute for Particle Physics (SCIPP)
charge-collection (CC) apparatus (also described below) before irradiation.
The sensors remained mounted to their individual daughter boards throughout irradiation and the followup 
tests, simplifying their handling and reducing uncontrolled annealing. 
Additionally, this allowed a reverse-bias voltage to be maintained across the sensor during irradiation.
The voltage was kept small (at the level of a few volts) to avoid possible damage of the devices 
from a large instantaneous charge during the spill.

Sensors were irradiated with beam provided by the End Station
Test Beam (ESTB) facility at the SLAC National Accelerator Laboratory.
Parameters of the beam provided by the ESTB facility are shown in
Table~\ref{tab:ESTB}. The beam was incident upon a series of 
tungsten radiators, as enumerated in Table~\ref{tab:setup}.
An initial 7mm-thick tungsten plate 
served to initiate the electromagnetic shower.
The small number of
radiation lengths of this initial radiator (2.0) permitted the
development of a small amount of divergence of the shower
relative to the straight-ahead beam direction without
significant development of the largely isotropic hadronic 
component of the shower.

\begin{table*}[h]
\begin{centering}
\caption{Parameters of the beam delivered by the
ESTB facility during the T-506 experiment.}
\label{tab:ESTB}
\vspace {5mm}
\begin{tabular}{cc}
Parameter  &  Value   \\ \hline
Energy     &  3.5-10.6 GeV \\
Repetition Rate  &  5-10 Hz \\
Charge per Pulse  &  150 pC \\
Spot Size (radius) & $\sim 1$ mm \\
\end{tabular}
\par\end{centering}
\end{table*}

This plate was followed by an
open length of approximately 55cm, which allowed a degree of
spreading of the shower before it impinged upon a second,
significantly thicker radiator (4.0 radiation lengths)
which was followed immediately by the sensor undergoing
irradiation. This was closely followed, in turn, by an
8.0 radiation-length radiator. Immediately surrounding the
sensor by tungsten radiators that both
initiated and absorbed the great majority of the electromagnetic
shower ensured that the sensor would be illuminated by a
flux of hadrons commensurate with that experienced by a calorimeter
sensor close to the maximum of a tungsten-induced shower.

\begin{table*}[h]
\begin{centering}
\caption{\label{tab:target_config} Elements encountered by
incident beam as a function of longitudinal position,
relative to the upstream face of the initial tungsten
radiator.}
\label{tab:setup}
\vspace {5mm}
\begin{tabular}{lcc}
Element  &  Longitudinal   & Accumulated \\ 
         &     Position (cm)        & Radiation Lengths \\ \hline
Front face radiator 1         &  $0.0$  & 0.0  \\
Back face radiator 1          &  $0.7$  & 2.0 \\
Front face radiator 2         & $55.0$  & 2.0  \\
Back face radiator 2          & $55.7$  & 4.0  \\
Front face radiator 3         & $56.3$  & 4.0  \\
Back face radiator 3          & $57.0$  & 6.0  \\
Sensor sample                 & $57.7$  & 6.0 \\
Front face radiator 4         & $59.0$  & 6.0  \\
Back face radiator 4          & $60.4$  & 10.0  \\
Front face radiator 5         & $61.0$  & 10.0  \\
Back face radiator 5          & $62.4$  & 14.0  \\
\end{tabular}
\par\end{centering}
\end{table*}

Although initiating the shower significantly upstream of the sensor
promoted a more even illumination of the sensor 
than would otherwise have been achieved, the half-width
of the resulting electron-positron fluence distribution
at the sensor plane was less than 0.5 cm. On the other hand,
the aperture of the CC apparatus (to be
described below) was of order 0.7 cm.. Thus, in order to 
ensure that the radiation dose was well understood over 
the region of exposure to the CC apparatus source,
it was necessary to achieve a uniform illumination over 
a region of approximately 1cm$^2$. This was done by
`rastering' the detector across the beam spot through
a range of 1cm in the directions both along
and transverse to the direction of the sensor's strips,
generating a region of approximately 1cm$^2$ over which 
the illumination was uniform to within $\pm 5$\%. 

\section{Dose Rates}

During the 120 Hz operation of the SLAC Linear Collider Light Source (LCLS),
5-10 Hz of beam was deflected by a pulsed kicker magnet into the End Station transfer line.
The LCLS beam was very stable with respect to both current and energy. Electronic
pickups and ion chambers measured the beam current and beam loss through the
transfer line aperture, ensuring that good transfer efficiency could be established
and maintained. The transfer efficiency was estimated to be ($95 \pm 5$)\%, although
for the highest energy beams delivered in the final days of T-506, the transfer line
experienced small but persistent beam loss; for this period, the transfer
efficiency was measured to be ($90 \pm 10$)\%. These transfer factors and their
uncertainties were taken into account in the estimation of dose rates through
the exposed sensors.

To calculate the dose rate through the sensor, it is necessary to determine
the `shower conversion factor' $\alpha$ that provides the mean fluence of minimum-ionizing 
particles (predominantly electrons and positrons), in particles per cm$^2$,
per incoming beam electron. This factor is dependent upon the radiator
configuration and incident beam energy, as well as the rastering pattern 
used to provide an even fluence across the sensor (as stated above,
the detector was translated continuously across the beam centerline
in a 1 cm$^2$ square pattern).

To estimate $\alpha$, the Electron-Gamma-Shower (EGS) Monte Carlo program~\cite{ref:EGS}
was used to simulate showers through the radiator configuration
and into the sensor. The configuration of Table~\ref{tab:setup}
was input to the EGS program, and a mean fluence profile (particles per
cm$^2$ through the sensor as a function of transverse distance from the nominal
beam trajectory) was accumulated by simulating the showers of 1000
incident electrons of a given energy. To simulate the rastering process,
the center of the simulated profile was then 
moved across the face of the sensor in 0.5mm steps, and an estimated mean fluence
per incident electron as a function of position on the sensor (again, relative to the nominal beam
trajectory) was calculated. This resulted in a mean fluence per incident electron
that was uniform to within a few percent 1mm or more inside of the edge of the rastering
region.
The value of $\alpha$ used for subsequent irradiation dose estimates was taken to be
the value found at the intersection of the nominal
beam trajectory with the sensor plane. The simulation was repeated for
various values of the incident electron energy, producing the values of
$\alpha$ shown in Table~\ref{tab:alpha}.

\begin{table*}[h]
\begin{centering}
\caption{Shower conversion factor $\alpha$, giving
the mean fluence at the sensor per incident
electron, as a function of electron energy. These
values include the effect of rastering over a 1 cm$^2$
area surrounding the nominal beam trajectory.
Also shown is the number of Rads per nC of delivered
charge, at the given energy, corresponding to the
given value of $\alpha$.}
\label{tab:alpha}
\vspace {5mm}
\begin{tabular}{lcc}
Electron      & Shower Conversion  & Dose per nC Delivered   \\
Energy (GeV)  & Factor $\alpha$    & Charge (kRad)           \\ \hline
2  &  2.1 & 0.34  \\
4  &  9.4 & 1.50  \\
6  & 16.5 & 2.64  \\
8  & 23.5 & 3.76  \\
10 & 30.2 & 4.83  \\
12 & 36.8 & 5.89  \\
\end{tabular}
\par\end{centering}
\end{table*}

To convert this number to Rads per nC of delivered charge, a mean
energy loss in silicon of 3.7 MeV/cm was assumed, leading to
a fluence-to-Rad conversion factor of 160 Rad per nC/cm$^2$.
It should be noted that, while this dose rate considers only
the contribution from electrons and positrons, these
two sources dominate the overall energy absorbed by the
sensor. In addition, the BeamCal dose-rate spec of 100 MRad
per year considered only the contribution from electrons
and positrons.

In order to accurately estimate the dose rates, it was also necessary to ensure that
the nominal beam trajectory passed through a well-known
and reproducible position on the sensors. A jig attached
to the downstream side of Radiator 3 (see Table~\ref{tab:setup})
positioned the daughter board carrying the sensor at
a fixed position relative to the radiator configuration.
Each sensor was mounted onto its own daughter board at 
a location reproducible to sub-millimeter accuracy. The
desired location of the nominal beam trajectory in the 
middle of the 1 cm$^2$ rastering pattern was then transfered
to the upstream face of Radiator 2, which was rigidly attached
to Radiator 3, using a mechanical metrology procedure. A Delrin
pin was attached to the upstream face of Radiator 2 at a
known displacement from the desired beam location, which
was then used to spindle a reticled phosphorescent screen.
The sensor/radiator assembly was then moved to the 
center of the rastering pattern, and with Radiator 1
removed, the beam was
steered until it hit the intended place on the
reticled screen. With the beam trajectory thus
established, Radiator 1 was replaced and two upstream
phosphorescent screens were placed in the beamline. 
The position of the beam on these screens was
recorded, establishing both the position and
angle of the properly steered beam. 

To confirm the adequacy of the dose-calibration simulation
(described above) and this alignment procedure, an 
in-situ measurement of the dose was made using a
radiation-sensing field-effect transistor (`RADFET')~\cite{ref:radfet}
positioned on a daughter board at the expected
position of the nominal beam trajectory at the
center of the rastering pattern. 
Beam was delivered in 150 pC pulses of 4.02 GeV
electrons; a total of 1160 pulses were directed
into the target over a period of four minutes,
during which the sensor was rastered quickly
through its 1 cm$^2$ pattern.
The RADFET was then read out, indicating
a total accumulated dose of 230 kRad, 
with an uncertainty of roughly 10\%. Making
use of the dose rate calibration of Table~\ref{tab:alpha},
interpolating to the exact incident energy of 4.02 GeV,
and taking into account the ($95 \pm 5$)\% transfer efficiency
of the ESTB beamline, leads to an expected dose of 250 kRad,
within the $\sim$10\% uncertainty of the RADFET measurement.

\section{Sensor Irradiation Levels}

As mentioned above, four types of sensors were studied:
p-type and n-type doped versions of
both magnetic Czochralski and float-zone crystals.
In what follows, we will use the notation `N' (`P')
for n-type (p-type) bulk sensors, and `F' (`C')
for float-zone (magnetic Czochralski) crystal technology.
Once a sensor was irradiated with the ESTB, it was placed
in a sub-freezing environment and not irradiated again.
Up to four sensors of each type were irradiated and
chilled until they could be brought back to the University
of California, Santa Cruz campus for the
post-irradiation CC measurement. In
addition, the sub-freezing environment was maintained
both during and after the CC measurement, so
that controlled annealing studies can eventually be done.

Table~\ref{tab:dose} displays the dose parameters of the 
irradiated sensors. The
$(95 \pm 5)$\% transfer line efficiency has been taken
into account in these estimates. The numeral following
the two letters in the sensor identifier refer to
an arbitrary ordering of sensors assigned during
the sensor selection. Sensors were held at between 0 and 5 C
during irradiation. With the exception of sensor
NC02, which was accidentally annealed for 5 hours at temperatures as high as
130 C, all sensors were transferred to a cold
(below -10 C) environment immediately after irradiation.
All four sensor types were exposed to dose rates
of approximately 5 and 20 MRad, while an NF sensor
received over 90 MRad and an NC sensor 220 MRad.
CC results for the 
irradiated sensors will be presented below.

\begin{table*}[h]
\begin{centering}
\caption{Dose parameters of the irradiated sensors. The 
$(95 \pm 5)$\% transfer line efficiency has been taken
into account in these estimates. While the NC02
sensor was irradiated at a temperature of 5 C,
it was accidentally annealed for approximately 5 hours
at temperatures as high as 130 C. Also included is the
minimum voltage for full depletion ($V_{FD}$) for
the sensors before irradiation.} 
\label{tab:dose}
\vspace {5mm}
\begin{tabular}{lccccc}
Sensor      & $V_{FD}$  & Irradiation      & Beam Energy   &   Delivered      &  Dose     \\
         &     & Temp. (C)  & (GeV)         &   Charge ($\mu$C) & (MRad)    \\ \hline \hline
PF05   & 190 & 0 & 5.88                & 2.00               & 5.13  \\
PF14   & 190 & 0 & 3.48                & 16.4               & 19.7  \\ \hline
PC10   & 660 & 0 & 5.88                & 1.99               & 5.12  \\
PC08   & 700 & 0 & (5.88, 4.11, 4.18)  & (3.82,3.33,3.29)   & 20.3  \\ \hline
NF01   &  90 & 0 & 4.18                & 2.30               & 3.68  \\
NF02   &  90 & 0 & 4.02                & 12.6               & 19.0  \\ 
NF07   & 100 & 5 & 8.20                & 23.6               & 91.4  \\ \hline
NC01   & 220 & 0 & 5.88                & 2.00               & 5.13  \\
NC10   & 220 & 0 & 3.48                & 15.1               & 18.0  \\
NC03   & 220 & 5 & 4.01                & 59.9               & 90.2  \\
NC02   & 220 & 5$^*$ & (10.60,8.20)        & (32.3,13.8)        & 220   \\  \hline
\end{tabular}
\par\end{centering}
\end{table*}

\section{Charge Collection Measurement}

The SCIPP CC
apparatus incorporates a $^{90}$Sr source that has a secondary $\beta$-decay
with an end-point energy of 2.28 MeV. These $\beta$ particles illuminate
the sensor under study, 64 channels of which are read out by the PMFE ASIC~\cite{ref:PMFE},
with a shaping time of 300 nsec. Whenever one of the 64 channels exceeds 
a pre-set, adjustable threshold, the time and duration of the excursion 
over threshold is recorded. In addition, the $\sim$250 Hz of $\beta$ particles that pass through
the sensor, and subsequently
enter a small (2mm horizontal by 7mm vertical) slit, trigger a
scintillator, and the time of excitation of the scintillator is also recorded.
If the slit is properly aligned with the read-out channels of
the sensor, and the sensor is efficient at the set read-out threshold,
a temporal coincidence between the scintillator pulse and
one of the read-out channels will be found in the data stream.

\begin{figure*}[h]
 \begin{centering}
  \includegraphics[scale=0.55]{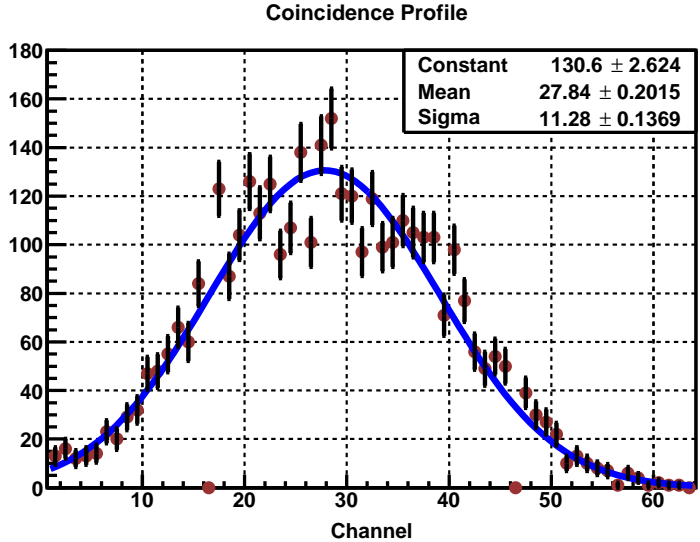}
 \par\end{centering}
 \caption{Sample profile of coincidences between the read-out
sensor channels and the trigger scintillator. The integral
of this distribution provides a count of the number of
$\beta$ particles triggering the scintillator that also
exceed the chosen PMFE threshold in one of the read-out channels.
\label{fig:coinc}
}
\end{figure*}

\begin{figure*}[h]
 \begin{centering}
  \includegraphics[scale=0.55]{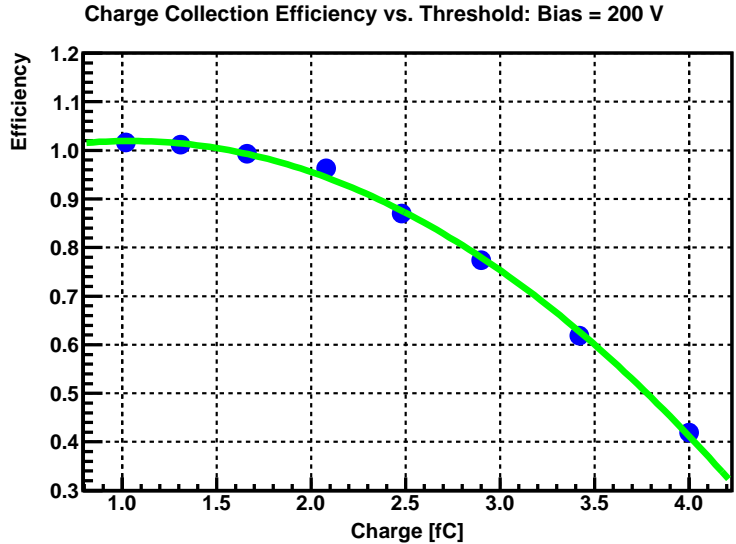}
 \par\end{centering}
 \caption{
Plot of efficiency vs. PMFE threshold setting for
one of the irradiated sensors. The abscissa has 
been converted from applied threshold voltage to
the amount of input PMFE charge that will exceed
the given threshold exactly 50\% of the time.
The point at which the curve
crosses the 50\% level
yields the median CC for the given bias level.
\label{fig:thresh_curve}
}
\end{figure*}

Figure~\ref{fig:coinc} shows a sample coincidence profile 
(histogram of the number of coincidences vs. channel
number) for a 150-second run at a given threshold and
reverse bias level for one of the irradiated sensors
(specifically, for the NC01 sensor after 5.1 MRad
of irradiation, applying a 300V reverse bias and a 130 mV threshold). 
The integral of the distribution yields an estimate
of the total number of coincidences found during
the run, which, when divided by the number of scintillator
firings (after a small correction for cosmic background events)
yields the median CC level at that threshold 
and bias level. This measurement can then be performed as
a function of threshold level, yielding the curve shown in
Figure~\ref{fig:thresh_curve}. For this plot, the abscissa
has been converted from voltage (the applied threshold
level) to fC (the PMFE input charge that will fire the threshold
with exactly 50\% efficiency) via a prior calibration step
involving measurement of the PMFE response to known 
values of injected charge. The point at which the curve
in Figure~\ref{fig:thresh_curve} crosses the 50\% level
yields the median CC for the given bias level.
In a prior study of sensors irradiated with hadrons, the SCIPP
apparatus gave median charge results consistent with that of 
other charge collection systems used to assess radiation damage in
that study~\cite{Hara}.

\section{Charge Collection Results}

The daughter boards
containing the irradiated sensors were designed
with connectors that allowed them to be attached to the
CC apparatus readout board without handling the sensors.
The median CC was measured as a function of reverse bias voltage for each sensor 
both before and after irradiation.

\begin{figure*}[h]
 \begin{centering}
  \includegraphics[scale=0.55]{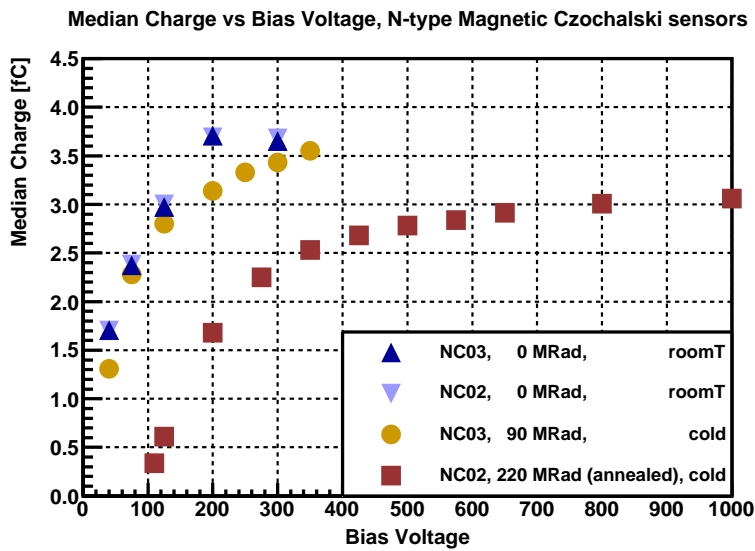}
 \par\end{centering}
 \caption{
High-dose CC results for the NC03 (90 MRad dose) and NC02 (220 MRad dose)
sensors.
\label{fig:NC_CCE}
}
\end{figure*}

The best performance was observed for the NC (n-type bulk
magnetic Czochralski) sensor type. For the exposures of 5.1 (NC01)
and 18.0 MRad (NC10), no difference in charge collection performance was observed
relative to the pre-irradiation studies of the NC01 and NC10 sensors.
In Figure~\ref{fig:NC_CCE} the median CC
both before and after irradiation is plotted for the NC03 (90 MRad dose)
and NC02 (220 MRad dose) sensors; it should be borne in mind, 
though, that the NC02 sensor experienced significant annealing before
the post-irradiation measurement was done. It is seen that,
while the depletion voltage increases significantly with dose, median CC
within 20\% of un-irradiated values is maintained
for doses above 200 MRad, although it may require annealing
to maintain efficiency at that level.

\begin{figure*}[h]
 \begin{centering}
  \includegraphics[scale=0.55]{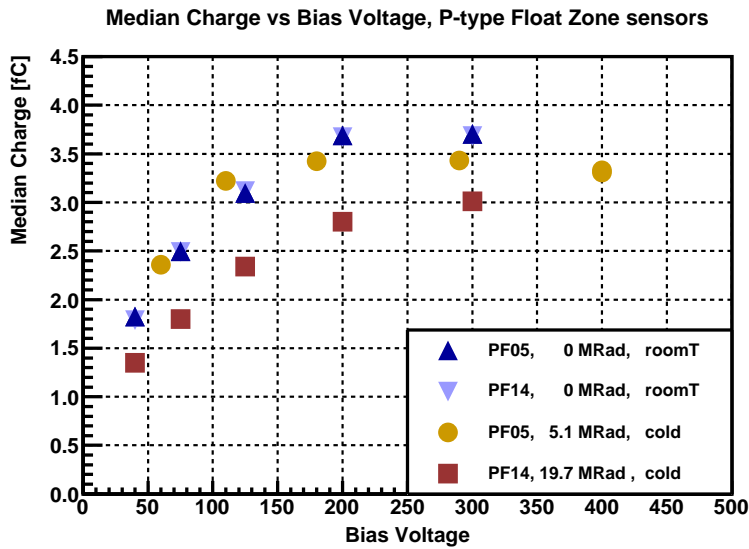}
 \par\end{centering}
 \caption{
CC results for the PF05 (5.1 MRad dose) and PF14 (19.7 MRad dose)
sensors.
\label{fig:PF_CCE}
}
\end{figure*}

\begin{figure*}[h]
 \begin{centering}
  \includegraphics[scale=0.55]{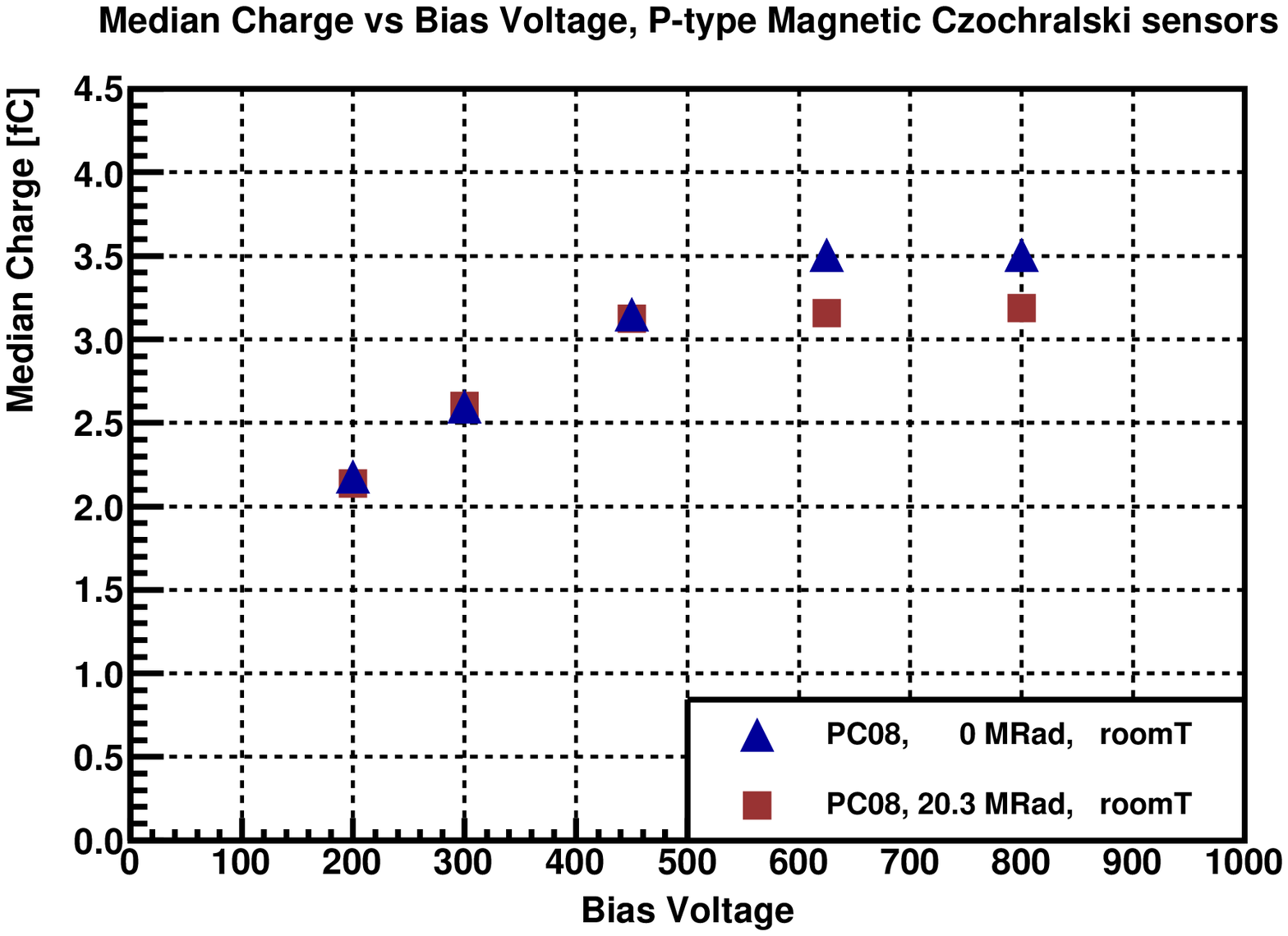}
 \par\end{centering}
 \caption{
CC results for the PC08 (20.3 MRad dose) sensor. The
PC10 sensor (5.1 MRad dose) suffered damage during
handling and did not give reliable results. 
\label{fig:PC_CCE}
}
\end{figure*}

\begin{figure*}[h]
 \begin{centering}
  \includegraphics[scale=0.55]{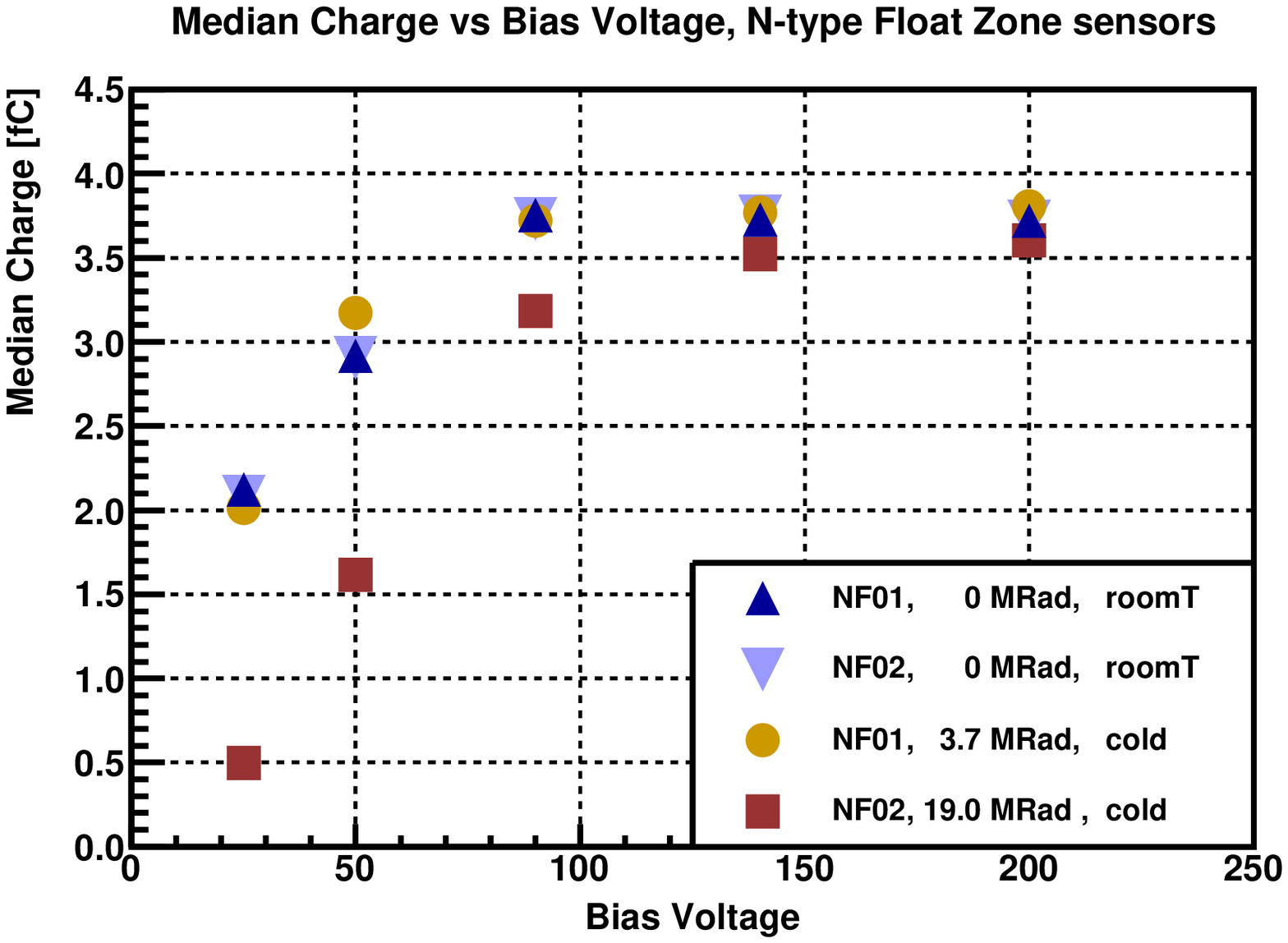}
 \par\end{centering}
 \caption{
CC results for the NF01 (3.7 MRad dose) and NF02 (19.0 MRad dose)
sensors.
\label{fig:NF_CCE}
}
\end{figure*}

Figures~\ref{fig:PF_CCE} through~\ref{fig:NF_CCE} show the results
for the remaining three sensor types (PF, PC and NF) for
irradiation levels up to approximately 20 MRad. Charge collection 
remains high for the PC and NF sensors
at this dose level, with the PF sensors showing 10-20\%
charge collection loss at 19.7 MRad. While this represents a dose of only
about 20\% of the expected annual dose for the most heavily-irradiated
sensors in the BeamCal instrument, it is possible that a period
of controlled annealing may restore some or all of the CC loss
for these sensors. An NF sensor (NF07) with a 91 MRad
exposure remains to be evaluated with the SCIPP CC apparatus.

Table~\ref{tab:cc_results} provides a table of maximum median collected
charge, both before and after irradiation, and median
charge loss due to irradiation. Not shown are results for the PC10 (damaged
during handling) and NF07 (still under study) sensors.

\begin{table*}[h]
\begin{centering}
\caption{Maximum median charge collection before and
after irradiation. While the NC02
sensor was irradiated at a temperature of 5 C,
it was accidentally annealed for approximately 5 hours
at temperatures as high as 130 C.}
\label{tab:cc_results}
\vspace {5mm}
\begin{tabular}{lcccc}
Sensor    & Dose    & Median CC Before  & Median CC After      &  Fractional      \\
         &   (MRad) &     Irradiation (fC) & Irradiation (fC)         & Loss (\%)    \\ \hline \hline
PF05   & 5.1  &  3.70  &  3.43  &   7     \\
PF14   & 20   &  3.68  &  3.01  &  18     \\
PC08   & 20   &  3.51  &  3.09  &  12     \\
NF01   & 3.7  &  3.76  &  3.81  &   0     \\ 
NF02   & 19   &  3.75  &  3.60  &   4     \\
NC01   & 5.1  &  3.71  &  3.80  &   0     \\
NC10   & 18   &  3.76  &  3.74  &   1     \\
NC03   & 90   &  3.68  &  3.55  &   4     \\
NC02   & 220  &  3.69  &  3.06  &  17$^*$ \\  \hline
\end{tabular}
\par\end{centering}
\end{table*}

\section{Summary and Conclusions}

We have explored the radiation tolerance of four different types of 
silicon diode sensors (n-type and p-type Float Zone and Magnetic Czochralski bulk sensors), 
exposing them to doses as high as 220 MRad
at the approximate maxima of tungsten-induced electromagnetic showers.
We have found all types to be radiation tolerant to 20 MRad, with the
n-type Czochralski sensors exhibiting less than a 20\% reduction 
in median collected charge for a dose in excess of 200 MRad. This
suggests the possibility of charge collection
sufficient for the operation of a calorimeter exposed to
hundreds of MRad, approaching the specification required for 
the most heavily irradiated sensors in the ILC BeamCal instrument. We plan to
follow through with IV and CV studies of the irradiated sensors, as well as
annealing studies on selected sensors.

\section{Acknowledgments}

We are grateful to Leszek Zawiejski, INP, Krakow for supplying us with the tungsten plates 
needed to form our radiator. We also would like to express our gratitude
to the SLAC Laboratory, and particularly the End Station Test Beam delivery
and support personnel, who made the run possible and successful.
Finally, we would like to thank our SCIPP colleague Hartmut Sadrozinski for
the numerous helpful discussions and guidance he provided us.

\section{Role of the Funding Source}

The work described in this article was supported by the United States Department of Energy,
DOE contract DE-AC02-7600515 (SLAC) and grant DE-FG02-04ER41286 (UCSC/SCIPP). The funding agency
played no role in the design, execution, interpretation, or
documentation of the work described herein.





\bibliographystyle{/usr/share/texmf-texlive/tex/latex/elsevier/model1-num-names}
\addcontentsline{toc}{section}{\refname}\bibliography{thesis-citations}




\end{document}